\documentclass[10pt, conference, letterpaper]{IEEEtran}
 \pdfoutput=1 
\usepackage{subfiles}
\usepackage{xspace}
\usepackage{graphicx}
\let\OLDthebibliography\thebibliography
\renewcommand\thebibliography[1]{
 \OLDthebibliography{#1}
 \setlength{\parskip}{0pt}
 \setlength{\itemsep}{0pt plus 0ex}
}
\usepackage[utf8]{inputenc}
\usepackage[english]{babel}
\newcommand{\papertitle}{WAKU-RLN-RELAY\xspace}
\newcommand{\waku}{WAKU\xspace}
\newcommand{\wakurelay}{WAKU-RELAY\xspace}

\newcommand{\ext}{\emptyset}
\newcommand{\proof}{\pi}
\newcommand{\intnul}{\phi}
\usepackage{tikz} % for IEEE copyright
\usepackage{textcomp}
\usepackage{hyperref}
\newcommand\copyrighttext{%
  \footnotesize \textcopyright 
  20xx IEEE. Personal use of this material is permitted. Permission from IEEE must be obtained for all other uses, in any current or future media, including reprinting/republishing this material for advertising or promotional purposes, creating new collective works, for resale or redistribution to servers or lists, or reuse of any copyrighted component of this work in other works.
  }
\newcommand\copyrightnotice{%
\begin{tikzpicture}[remember picture,overlay]
\node[anchor=south,yshift=10pt] at (current page.south) {\fbox{\parbox{\dimexpr\textwidth-\fboxsep-\fboxrule\relax}{\copyrighttext}}};
\end{tikzpicture}%
}
\begin{document}
\IEEEoverridecommandlockouts % to display thanks https://tex.stackexchange.com/questions/53546/thanks-wont-appear-in-ieeetran
\title{ Privacy-Preserving Spam-Protected Gossip-Based Routing}

\author{\IEEEauthorblockN{Sanaz Taheri-Boshrooyeh\IEEEauthorrefmark{1}\IEEEauthorrefmark{2},
Oskar Thor\'en\IEEEauthorrefmark{1}\IEEEauthorrefmark{2}, 
Barry Whitehat\IEEEauthorrefmark{3},
Wei Jie Koh\IEEEauthorrefmark{4},
Onur Kilic\IEEEauthorrefmark{5}, and
Kobi Gurkan\IEEEauthorrefmark{6}}
\IEEEauthorblockA{
\IEEEauthorrefmark{1}Vac Research and Development,
\IEEEauthorrefmark{2}Status Research and Development, Singapore,
\IEEEauthorrefmark{3}Unaffiliated,\\
\IEEEauthorrefmark{4}Independent,
\IEEEauthorrefmark{5}Unaffiliated,
\IEEEauthorrefmark{6}cLabs 
}
\IEEEauthorblockA{
% \IEEEauthorrefmark{1}
sanaz@status.im,
% \IEEEauthorrefmark{2}
oskar@status.im,
% \IEEEauthorrefmark{3}
barrywhitehat@protonmail.com,\\
% \IEEEauthorrefmark{4}
contact@kohweijie.com,
% \IEEEauthorrefmark{5}
onurkilic@protonmail.com,
% \IEEEauthorrefmark{6}
me@kobi.one
}
}
\maketitle
\copyrightnotice

\begin{abstract}
% abstract of fewer than 200 words
\papertitle is an anonymous peer-to-peer gossip-based routing protocol that features a privacy-preserving spam-protection with cryptographically guaranteed economic incentives. 
While being an anonymous routing protocol where routed messages are not attributable to their origin, it allows global identification and removal of spammers. 
It addresses the performance and privacy issues of its counterparts including proof-of-work and reputation-based schemes.
Its light computational overhead makes it suitable for resource-limited environments. 
The spam protection works by limiting the messaging rate of each network participant where rate violation results in financial punishment. 
We deploy the novel construct of rate-limiting nullifier to enforce the message rate limit. 
We provide a proof-of-concept implementation of \papertitle to prove the efficiency and feasibility of our solution.
\end{abstract}

\begin{IEEEkeywords}
P2P, Spam Protection, Messaging, zkSNARKs, Zero-Knowledge, Anonymity, Routing, Pub/Sub, Gossipsub
\end{IEEEkeywords}
\section{Introduction}
\waku \cite{vacrfc}
is a family of peer-to-peer (p2p) protocols that enable anonymous and censorship-resistant communication over a network of heterogeneous peers including resource-restricted devices. It features \wakurelay protocol \cite{vacrfc}
which is an anonymous gossip-based Pub/Sub protocol. Peers congregate around topics and can send messages to topics rather than individuals. Messages are routed using the gossip protocol.  Receiver anonymity results from the gossip nature of the routing protocol. Sender anonymity is protected by anonymizing protocol messages i.e., removing personally identifiable information (PII) that binds a message to its owner e.g., IP address and digital signatures.

Being anonymous and an open messaging network, \wakurelay is prone to spam activities where spammers send bulk messages, exhaust the resources of the entire p2p network and cause a denial of service attack. Global identification and removal of spammers are required to protect network resources and liveness. However, as messages are not attributable to their origin,  conventional spam-prevention techniques like IP blocking are not effective. 

None of the state-of-the-art p2p spam protection techniques i.e.,  \textit{Proof of Work} (PoW) \cite{dwork1992pricing}  deployed by Whisper\footnote{https://eips.ethereum.org/EIPS/eip-627}

and \textit{Peer scoring}  adopted by Libp2p GossipSub \cite{vyzovitis2020gossipsub} provide global spam-protection. Moreover, PoW is computationally expensive hence not suitable for resource-constrained devices. The peer scoring is also prone to censorship and inexpensive attacks where millions of bots can be deployed to send bulk messages.
\begin{figure}[h]
\centering
\includegraphics[scale=0.15, trim={2cm 15cm 3cm 3cm}]{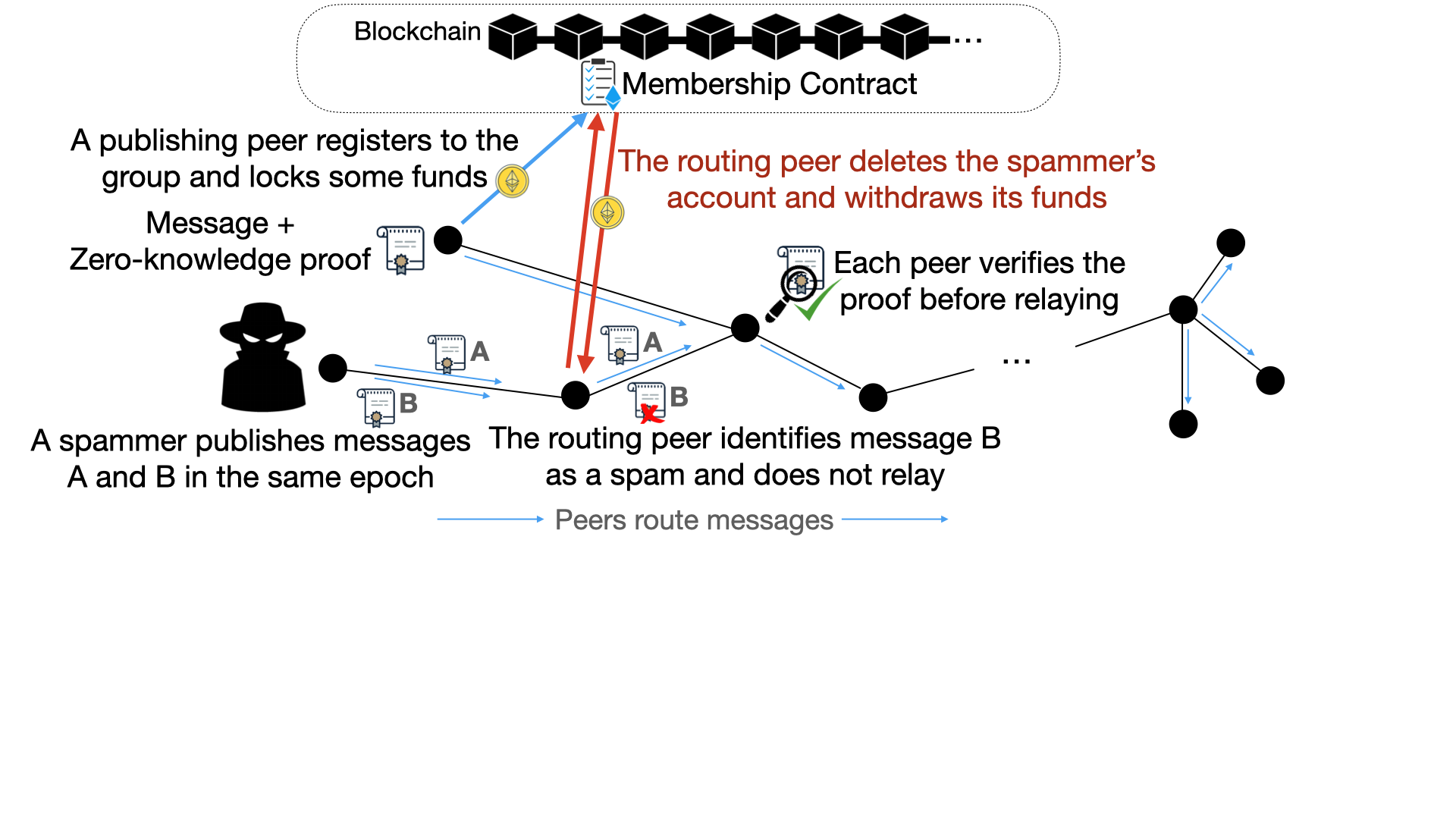}
    \caption{\papertitle Overview.}
    \label{fig:rln-relay-overview}
\end{figure}

\papertitle deals with the aforementioned issues. It extends \wakurelay with the novel construct of Rate Limiting Nullifiers (RLN) to enable an anonymous spam-protected p2p gossip-based Pub/Sub routing protocol which 1) is computationally efficient and suitable for resource-restricted devices 
2) preserves user anonymity 3) controls spammers globally 4) mitigates Sybil attack  5) has built-in economic incentives where spammers are financially punished and those who find spammers are rewarded.

\section{Preliminaries: Rate Limiting Nullifier}
RLN \cite{semaphore-slash} is a zero-knowledge and rate-limited signaling framework in which each user can only send $1$ message for each \textit{External Nullifier}. External nullifier is an application-dependent value and resembles a voting booth.
The group of authorized users is represented by a Merkle tree called membership tree whose leaves are members public keys $pk$. The corresponding secret keys $sk$ are known to their owning users. 
% $sk$ is a random element of a finite field whereas $pk$ 
$pk$s are cryptographic hash of $sk$s. The membership tree is stored in a smart contract deployed on the Ethereum blockchain. Users need to stake some ether in the contract to join the group. User removal is done by passing a member's secret key to the contract. 
Subsequently, a portion of the staked fund of the deleted member is burnt and a portion is given to whoever does deletion.

Only group members can signal. Users use zero-knowledge proof (specifically Zero-Knowledge Succinct Non-Interactive Argument of Knowledge \cite{groth2016size}) to prove their membership to the group without disclosing which members they are.  In a nutshell, zero-knowledge proof systems enable one party to prove the possession of certain information to a verifier without revealing other information.
Each RLN signal consists of the following parts $(m, \ext, \intnul, [sk], \proof)$. $m$ is the message. 
$\ext$ is the external nullifier. $\intnul=H(H(sk,\ext))$ is  \textit{internal nullifier} which is member's unique fingerprint for that external nullifier. $[sk]$ is a Shamir  secret share \cite{shamir1979share} of $sk$ deterministically derived from $m$, $sk$ and $\ext$. Two of such shares allow the reconstruction of $sk$.

$\proof$  is a zkSNARK proof asserting that $sk$ belongs to the membership tree as well as $[sk]$ and $\phi$ are correctly calculated. 

A signal is deemed valid if its $\proof$ field gets verified by the verification algorithm supplied by the underlying zkSNARK proof system. When a user creates two different messages for the same external nullifier, its signals end up with identical internal nullifiers which signify double-signaling. The two signals will each reveal a distinct share of the user's secret key by which one can reconstruct the $sk$ and remove the user.

\section{\papertitle Construction }
In this section, we explain the end-to-end integration of RLN in \papertitle and our design choices to achieve a cost-efficient and scalable solution that suites resource-limited devices. 

In \papertitle, a smart contract is deployed on the public Ethereum blockchain which acts merely as a registry keeping an ordered list of users public keys, unlike the initially proposed RLN construction which maintains the membership tree on the smart contract. The Merkle tree is maintained off-chain by individual peers. This design choice enables constant complexity registration and deletion operations (as opposed to logarithmic complexity in on-chain tree storage) hence optimizing gas consumption by an order of magnitude.

In contrast to RLN where signals are kept on the contract, in \papertitle,  messages are stored off-chain and distributed via p2p routing protocol of \wakurelay. Thus, we achieve higher massage propagation speed as opposed to the on-chain case where messages should be mined before being visible to the network. Moreover, we save our users the gas price that they have to otherwise pay to insert their messages to the contract.

An overview of \papertitle is provided in Figure \ref{fig:rln-relay-overview}. Peers that belong to the same GossipSub layer i.e., subscribed to the same topic form an RLN group. Peers route messages that belong to peers registered in the RLN group.

\textbf{Registration:} A peer submits a transaction to the membership contract embodying its public key $pk$ as well as $v$ amount of Eth to be staked on the contract.

\textbf{External nullifier}
We use $epoch$ as the external nullifier. $epoch$ is defined as the number of $T$ seconds that elapsed since the Unix epoch. 
Peers monitor the current $epoch$ locally and are allowed to publish one message per $epoch$.

\textbf{Publishing:}
Each peer is allowed to send one message $m$ per $epoch$. Message publishing in the network is the same as the RLN framework. The message owner, attaches the nullifiers, together with a  share of its secret  key, and the zero-knowledge proof part to the message. The peer wraps this data inside a GossipSub message and submits it to the network as in the normal routing protocol.

\textbf{Group Synchronization}: Publishing peers must always stay in sync with the latest state of the group. Otherwise,  by making proof of membership to an old version of the membership tree they can risk exposing the index of their public key in the tree hence compromising their anonymity. 
Upon member update, the membership contract emits update events by listening to which peers can  update their local trees.

\textbf{Routing and Slashing:} \label{sec:routing_slashing}
A routing peer follows the regular routing protocol of \wakurelay i.e., GossipSub protocol \cite{vacrfc} and additionally does the verification steps of the RLN framework. 
That is, it verifies the RLN proof and discards messages with invalid proof. The routing peer also  validates the epoch of the incoming message against its local epoch to see if their difference exceeds a threshold  $Thr$  in which case the message is considered invalid and gets dropped ($Thr=\frac{D}{T}$ where $D$ is the maximum network delay and $T$ is the length of the $epoch$). Epoch validation prevents a newly registered peer from spamming the system by messaging for all the past epochs. 
Next, the internal nullifier is verified to spot spam activity i.e., double-signaling. To do this, each routing peer locally keeps a record of the secret key share $[sk]$ and the internal nullifier $\phi$ of all of its incoming messages for the past $Thr$ epochs. This list is called a nullifier map. The routing peer checks every new message against this list to spot spam messages i.e., messages with identical internal nullifiers. 
Note that the nullifier map suffices to hold messages the belong to the last $Thr$ epochs because older messages are considered invalid by default. 
The message gets relayed if all checks pass.

\section{Security and Performance Analysis}
\textbf{Performance:} 
The proof of concept implementation of \papertitle is available in \cite{nim-waku,vacrfc} which uses RLN library \cite{rlnlib}. 
Generating membership proof to a group size of $2^{32}$ takes $\approx0.5$s on an iPhone 8 \cite{rlnlib}. Proof verification run time is constant and takes $\approx30$ms.
Each peer persists a $32$B public and secret keys and a $\approx3.89$MB prover key \cite{rlnlib}.
A membership tree with depth $20$ requires $67$MB storage which can be optimized to $0.128$KB using \cite{storage-efficient}.

\textbf{Security:} \papertitle achieves global spam protection while preserving user anonymity, namely peers 1) do not  disclose any piece of PII in any phase 2) prove their compliance with the messaging rate without leaving any trace to their public keys. Sybil attack is also mitigated by making registration expensive. Economic incentives are guaranteed cryptographically via secret sharing.

\bibliographystyle{IEEEtran}
\bibliography{refs}

\end{document}